\documentclass[letterpaper, 10 pt, conference]{ieeeconf}  

\IEEEoverridecommandlockouts                              
\usepackage{graphics} 
\usepackage{epsfig} 
\usepackage{mathptmx} 
\usepackage{times} 
\usepackage{amsmath} 
\usepackage{amssymb}  

\usepackage{tikzpagenodes}

\usepackage{epstopdf}
\usepackage{url} 

\usepackage{ar}
\usepackage{xfrac}
\newcommand\Rey{\mbox{\textit{Re}}}  
\newcommand\St{\mbox{\textit{St}}}  

\title{\LARGE \bf
A Novel Degree of Freedom in Flapping Wings Shows Promise for a Dual Aerial/Aquatic Vehicle Propulsor*
}

\author{Jacob S. Izraelevitz$^{1}$ and Michael S. Triantafyllou$^{2}$
\thanks{*This research was conducted with Government support under and awarded by DoD, Air Force Office of Scientific Research, National Defense Science and Engineering Graduate (NDSEG) Fellowship, 32 CFR 168a. Futhermore, this work was also supported by the CENSAM program within the Singapore-MIT Alliance for Research and Technology, and the MIT Sea Grant Program (contract number NA10OAR410086). }
\thanks{$^{1}$Jacob Izraelevitz is a Ph.D. Candidate in Mechanical Engineering at Massachusetts Institute of Technology
        {\tt\small jsi@mit.edu}}%
\thanks{$^{2}$Michael Triantafyllou is the William I. Koch Professor of Marine Technology and Professor of Mechanical Engineering at Massachusetts Institute of Technology
        {\tt\small mistetri@mit.edu}}%
}

\begin{document}

\maketitle
 \begin{tikzpicture}[remember picture,overlay]
    \node[text width = 7in, align=justify] at ([yshift=1em]current page text area.north) {\copyright 2015 IEEE. Personal use of this material is permitted. Permission from IEEE must be obtained for all other uses, in any current or future media, including reprinting/republishing this material for advertising or promotional purposes,  creating new collective works, for resale or redistribution to servers or lists, or reuse of any copyrighted component of this work in other works. 2015 IEEE International Conference on Robotics and Automation (ICRA) \bf DOI: 10.1109/ICRA.2015.7140015};
  \end{tikzpicture}

\thispagestyle{empty}
\pagestyle{empty}

\begin{abstract}

Ocean sampling for highly temporal phenomena, such as harmful algal blooms, necessitates a vehicle capable of fast aerial travel interspersed with an aquatic means of acquiring \emph{in-situ} measurements. Vehicle platforms with this capability have yet to be widely adopted by the oceanographic community. Several animal examples successfully make this aerial/aquatic transition using a flapping foil actuator, offering an existence proof for a viable vehicle design (Fig. \protect\ref{fig:CADrender}). We discuss a preliminary realization of a flapping wing actuation system for use in both air and water. The wing employs an active in-line motion degree of freedom to generate the large force envelope necessary for propulsion in both fluid media.

\end{abstract}

\section{INTRODUCTION}

\subsection{Ocean Sampling}

Data collection in the ocean is a challenging sparse-sampling problem; the shear volume of the ocean and range of spacial scales virtually guarantees that the phenomena of interest will be undersampled. Biological processes, in particular, are best measured by water sampling techniques rather than indirect methods, precluding the use of satellite systems. The scientific community's understanding of these phenomena is therefore limited to sensor data that give only a local view of the global problem, whether located on buoys, towed sensor arrays, or underwater vehicles.

Ocean behaviors with fast temporal dynamics are especially difficult to measure in this fashion, as the measurement of interest can vary faster than our ability to deploy sensors. Harmful algal blooms (HABs) are a primary example. HABs are highly localized, transient events that are virtually impossible to densely sample before the dynamics change without the capability for \emph{in-situ} sampling and processing \cite{babin2005new}. 

\begin{figure}[h]
\begin{centering}
\includegraphics[width=3.25in]{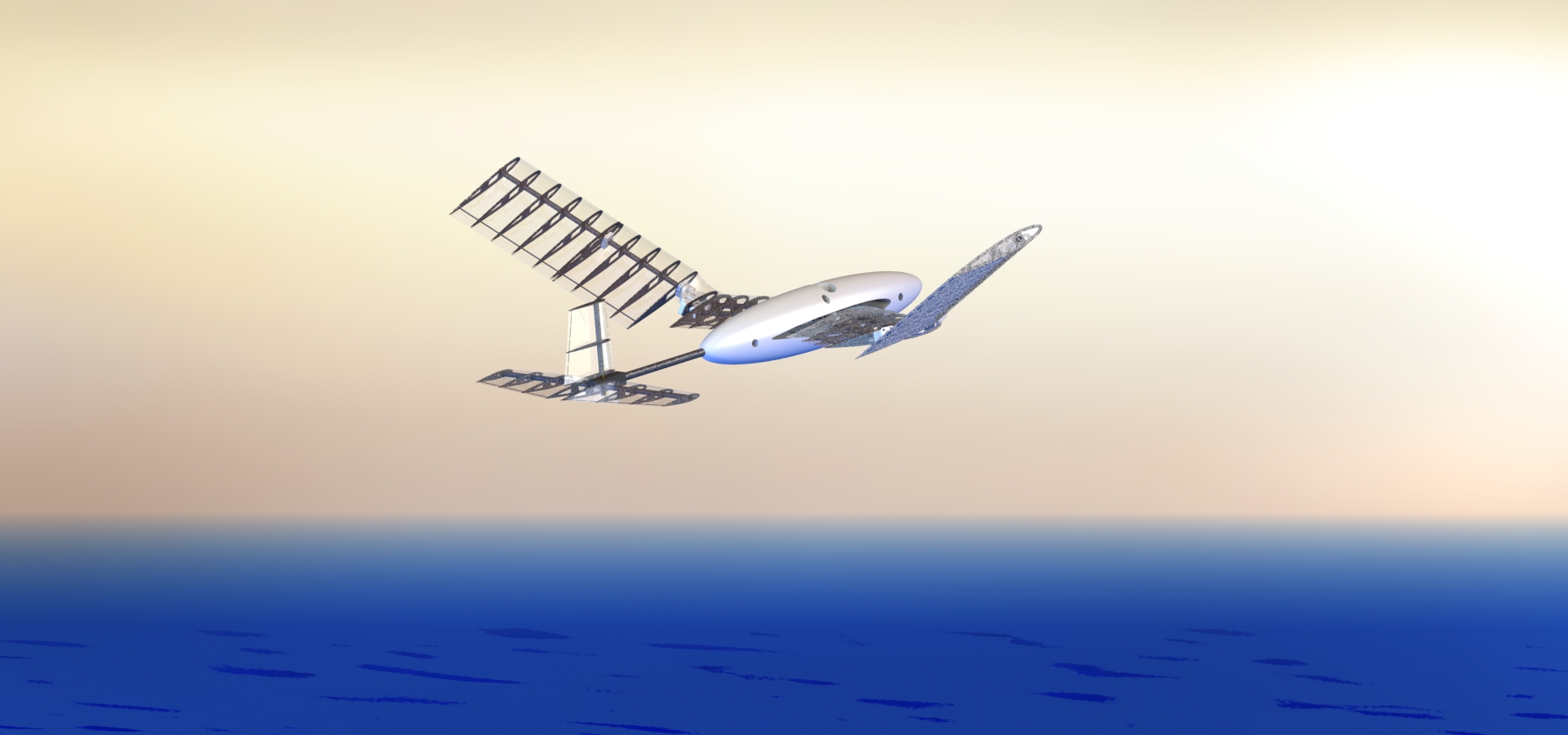}
\par\end{centering}
\caption{\emph{Dual Aerial/Aquatic Flapping Vehicle Concept} - Concept vehicle using in-line motion, an additional degree of freedom on a flapping wing, to provide the force necessary for propulsion in two fluids. \label{fig:CADrender}}
\end{figure}

In general, sampling systems are designed to optimize energy constraints for a given sampling area. However, if capturing transient dynamics is more important than the energy cost, then the limiting factor is the speed of the vehicle. This explains why aerial imaging techniques have had success in finding and measuring algal blooms \cite{sellner2003harmful}. The fluid physics simply favor fast travel in air, yet imagery cannot replace direct measurements for microbe identification or nutrient concentration. The ideal algal bloom sampling platform would use aerial imagery to find and characterize the bloom, while still retaining the ability to collect water samples. Our collaborator's previous attempts at HAB monitoring, coordinating the use of quadrotor imagery with surface vehicle measurements \cite{leighton2013system}, offer one such example of a dual aerial/aquatic approach.

\subsection{Transitional Vehicles}

To simplify the logistics, performing these measurements with a single aerial/aquatic vehicle platform is preferable. However, practical sampling vehicles of this form do not yet exist, despite early and ongoing development attempts. The history of manned aerial/aquatic vehicles dates back to several patents over a hundred years, although only Reid's prototype \cite{reid1963flying} was ever marginally functional. More recent unmanned efforts, all still under active development, include submarine-launched UAVs \cite{parry2013navy}, waterproof quadrotors \cite{drews2014hybrid}, and winged vehicles with jet-propelled takeoff \cite{siddall2014launching}. 

The primary actuation difficulties in creating such a vehicle, despite the identical governing equations in the two fluid media, can be summarized as follows:
\begin{itemize}
\item Large static lifting surfaces for weight support are unnecessary underwater, adding drag.
\item Underwater vehicles can offset extra weight with buoyant materials, so the design driver is often pressure hull integrity rather than vehicle mass.
\item The larger fluid density underwater generally leads to slower timescales in vehicle dynamics (given dynamic pressure equivalence and added mass).
\end{itemize}
In spite of these challenges, several biological examples prove that aerial/aquatic vehicles are indeed possible. Murres, puffins, and other auks both swim and fly using the same propulsor. Developing a transitional aerial/aquatic actuator resembling a flapping wing may unlock aspects to the problem not seen through traditional design techniques \cite{lock2014impact}.
\begin{figure}[t]
\begin{centering}
\includegraphics[scale=1.0]{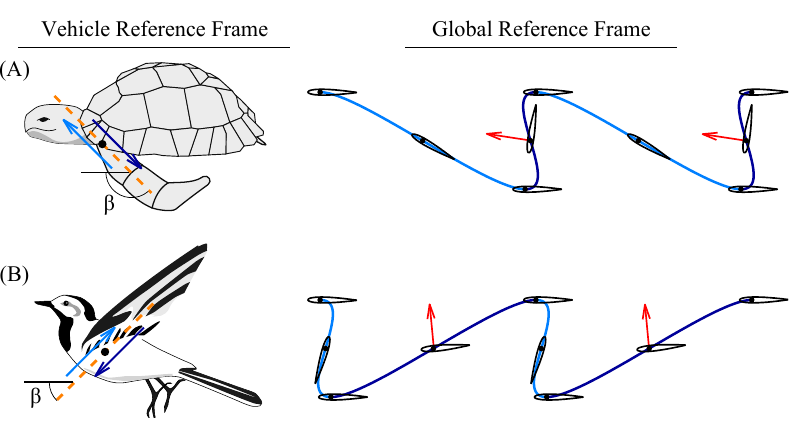}
\par\end{centering}
\caption{\emph{Comparison of Backwards In-line Downstroke (A), and Forwards In-line Downstroke (B)} - Various biological examples are able to change the direction of fluid force, denoted in red, by changing the stroke angle $\beta$ relative to oncoming flow. Figure modified from previous work in \protect\cite{izraelevitz2014adding}. \label{fig:IntroFig}}
\end{figure}

\section{BACKGROUND}

\subsection{In-line Motion}

Previous work in \cite{izraelevitz2014adding} investigates a highly improved range of force performance from a flapping wing using in-line motion, quantified through the stroke angle $\beta$. In-line motion \cite{licht2010line} is defined as a flapping motion where the foil is moved upstream or downstream in addition to across the flow (Fig. \ref{fig:IntroFig}). These types of trajectories are highly asymmetric.

In-line motion trajectories can either create thrust without an oscillating lifting force, similar to the flapping performance of turtles \cite{davenport1984comparison}, or create large lifting forces analogous to birds flying at slow speeds \cite{tobalske1996flight}. Coefficients of lift as high as $C_y=4$ can be achieved at the expense of expended power by the flapping actuator \cite{izraelevitz2014adding}. In addition, model-based optimization of flapping trajectory fluid force can mitigate unwanted variations in the fluid force \cite{izraelevitz2014adding}. 

Changing the degree of in-line motion could therefore be used as a means of force control for a flapping wing, and perhaps specifically aid in the aerial/aquatic dual design of a vehicle. Rather than relying on the static parameters of the airfoil to perform adequately in both air and water, the flapping trajectory could instead be modified.

\subsection{Flapping Foil Robots}

Flapping foil robots have been successfully designed to either perform impressive maneuvers underwater or provide weight support in air. Labriform aquatic swimming robots are often designed with kinematics that increasingly mimic turtle swimming. The Madeline aquatic tetrapod \cite{long2006four} and University of Tallinn U-CAT \cite{salumae2014design} employ only a single pitching actuator on each fin, but still maintain suitable vehicle control authority. The MIT Robotic Turtle \cite{licht2008biomimetic} and NTU robotic turtle \cite{low2007modular} combine a pitching and a flapping actuator. Finally, the RT-I incorporates a full four DOF in each fin to allow for both terrestrial walking and three modes of swimming \cite{kawamura2010design}. One of the presented modes uses in-line motion in a drag-based rowing motion, providing a starting point for developing a versatile actuator for use in an aerial/aquatic vehicle.

Many aerial flapping robots have also been developed that provide a suitable baseline of capability. In general, the flapping action is limited to a single degree of freedom, and the pitching action is achieved using the passive dynamics of a twisting wing. The U-Maryland Ornithopter \cite{grauer2009inertial} and MIT Kestrel \cite{jackowski2009design} are prime examples of this wing geometry using the hobbyist designs of Sean Kinkade \cite{kinkade2001ornithopter} as a template. The Festo SmartBird uses an active twisting mechanism and reports a greater measure of flight efficiency \cite{send2012artificial}. Numerous smaller micro-aerial vehicles (MAVs) have also been developed that operate in a quasi-hover regime and employ passive twisting mechanics, such as the Harvard RoboBee \cite{wood2008first}, the University of Delft DelFly II \cite{lentink2010scalable}, and the Nano Hummingbird of AeronVironment \cite{keennon2012development}.

The most analogous wing designs come from Lock et al. \cite{lock2014impact}, who are also investigating an aerial/aquatic flapping foil vehicle. Their strategy consists of a symmetric flapping trajectory and variable wingspan based off biological data from the common guillemot \emph{Uria Aalge}.

The general lesson from these vehicle examples is that adding degrees of freedom can increase performance: either efficiency, maneuverability, or travel modalities. However, while biological examples often have an unlimited number of degrees of freedom, an engineered implementation has to balance the performance with the cost of increased complexity of the design.  One therefore has to strongly justify whether additional degrees of freedom are necessary for the stated vehicle function. Accordingly, in-line motion represents an opportunity in the field, as the dramatic increase in vehicle functionality comes at a moderate cost in added complexity. In-line motion greatly widens the forcing capabilities of the actuator \cite{izraelevitz2014adding}, which appears necessary for a dual aerial/aquatic flapping wing.

\section{WING DESIGN}
\subsection{Design Concept}
\begin{figure}[t]
\begin{centering}
\includegraphics[scale=1.0]{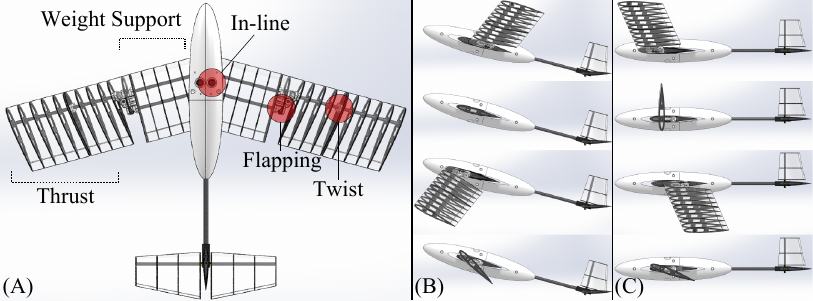}
\par\end{centering}
\caption{\emph{Wing Design Concept} - Shown from above in (A), in aerial configuration (B) and in aquatic configuration (C). Each image (top to bottom) taken at one-quarter of the flapping period, starting with the beginning of the downstroke. Figure modified from previous work in \protect\cite{izraelevitz2014model}. \label{fig:PrelimCAD}}
\end{figure}
The schematic in Fig. \ref{fig:PrelimCAD} illustrates our wing design concept of a theoretical vehicle. The in-line motion is actuated from the shoulder, shown in Fig. \ref{fig:PrelimCAD}A as a variable wing sweepback. The wing pitch and flapping motions are actuated only on the outer half of the wingspan. This nicely separates the two requirements of the wing - weight support and thrust generation. 

In air (Fig. \ref{fig:PrelimCAD}B), the vehicle flies at a net angle of attack, activating the wing area from the fuselage to midspan. Forwards in-line motion during the downstroke, biased by the angle of attack of the body, boosts the lift of both portions of the wing. The flapping of the outer wing, timed appropriately with the active pitching, provides the thrust.

Underwater (Fig. \ref{fig:PrelimCAD}C), the wing area from fuselage to midspan is deactivated by setting the angle of attack of the whole vehicle to zero. The outer wing still provides the thrust, following a turtle-like trajectory with strong backwards in-line motion during the downstroke.

\subsection{Experimental Prototype \label{proto}} 
\begin{figure}[t]
\begin{centering}
\includegraphics[scale=1.0]{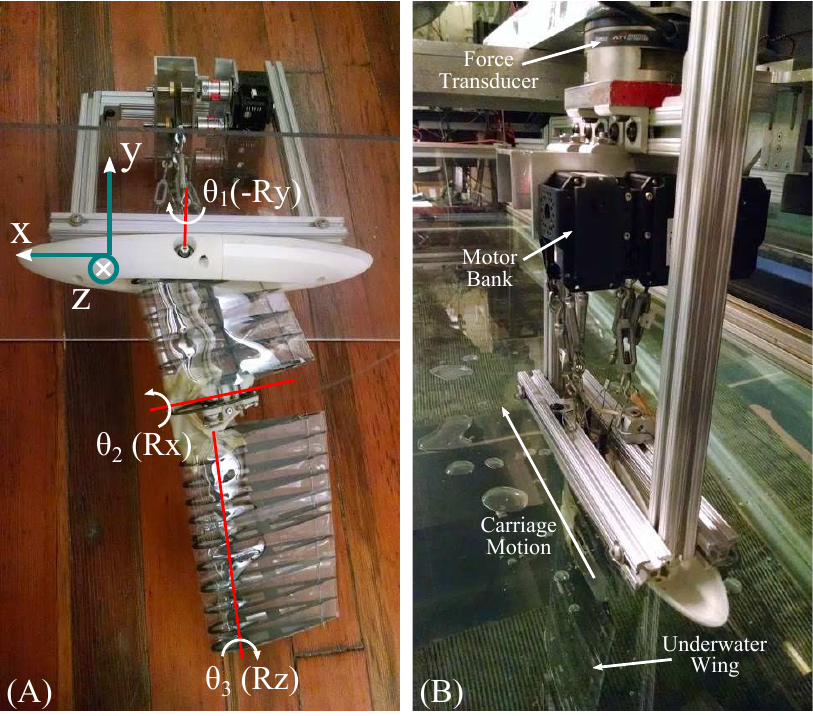}
\par\end{centering}
\caption{\emph{Towing Experiment} -  (A) Wing planform, cabling mechanism, and coordinate system. (B) Experiment to validate flapping scheme, where the wing is actuated and towed down the MIT small towing tank with a six axis force transducer. \label{fig:TowingExperiment}}
\end{figure}

For the purposes of validating the force performance of the in-line flapping concept, we built a first iteration half-vehicle model for experimental testing of the wing in the MIT Small Towing Tank (2.4 m long by 0.75 m width by 0.75 m depth), as illustrated in Fig. \ref{fig:TowingExperiment}. While the final vehicle is to be used for both aerial and aquatic travel, this first round of experimental testing was performed only on the wing geometry and only in water, similar to the design methodology employed by \cite{lock2014impact}. 

Surprisingly, both the aerial and aquatic modes of travel can be performed under nearly identical Reynolds number conditions $\Rey = Uc/\nu$, greatly simplifying the comparison of the fluid dynamics. Air has a greater kinematic viscosity ($\nu_\text{air}/\nu_\text{water}\approxeq12.5$ \cite{hoerner1965fluid}) that normalizes the increased aerial travel velocity $U$. We therefore non-dimensionalize the measured force to allow for evaluation in both fluid media given the water experiments, informing future vehicle design work in both air and water.

Given the compromise between our existing tank towing speed limits and force transducer measurement range, the aquatic velocity $U$ is fixed at $0.2$ m/s, giving us a Reynolds number of $\Rey = 30,000$, i.e. fast laminar flow. This corresponds to an aerial velocity of $2.5$ m/s. However, most of relevant fluid dynamics in this regime are only weakly dependent on Reynolds number (the primary exception being small skin drag that is dwarfed by pressure and induced drag), so we expect the resulting force coefficients to remain valid for the Reynolds number range $3,000 \leq \Rey \leq 300,000$.

The use of a half-vehicle model for validation assumes that the body of the vehicle will remain roughly stationary during flapping. This assumption breaks down during symmetric flapping, as the generation of thrust creates strong oscillating lift forces. However, for in-line motion flapping, the instantaneous direction of force is better aligned with the direction of travel, supporting the use of a stationary model.
As illustrated in Fig. \ref{fig:TowingExperiment}, the planform of the experimental prototype is rectangular, with the chord $c$ of 152 mm (6 in) and semispan $s$ three times the chord. The prototype  consists of three joints, located at coordinates easily non-dimensionalized by the chord. 
\begin{enumerate}
\item The in-line motion joint $\theta_1$ rotates about the negative y-axis, and is located at the wing root ($s_1=0$). The chordwise location of the joint is $c_1=0.375c$.
\item The flapping joint $\theta_2$ rotates about the x-axis, located at spanwise location $s_{2}=c$.
\item The pitching $\theta_3$ rotates about the z-axis at quarter chord, root located at spanwise coordinate $s_{3}=1.25c$.  
\end{enumerate}
The location of these joints was chosen by manipulation of the changing flow angle over the course of the wingspan during flapping (Fig. \ref{fig:FlowAngle}). Unlike symmetric flapping without in-line motion, the necessary pitch distribution over the wingspan is highly nonlinear. As such, the wing is required to dynamically pitch over a small area, necessitating a planform gap, sliding surface, or highly flexible material to maintain function. A suitable design compromise, allowing for over $\pm90^\circ$ of pitch, was a gap in the wing planform near the trailing edge, and a piece of latex to streamline the leading edge, covering the internal cabling.

The wing is built of carbon-fiber NACA0013 ribs covered with 0.5 mm thick mylar skin, with the inside of the wing intentionally flooded. The flooded wing, as opposed to a solid wing, is an imperative design requirement for aerial/aquatic use - as a solid wing must be made of a highly lightweight material for aerial flight, but adds tremendous unwanted buoyancy underwater. A symmetric NACA profile was again chosen for easy comparison with the flapping foil literature \cite{hover2004effect,techet2008propulsive,triantafyllou1991wake}, but other foil profiles would likely have better performance and is a focus for future work. Rib spacing is 28 mm, close enough to prevent buckling of the foil cross-section during untowed flapping (maximum loading), as validated through high-speed camera footage. 

The wing is mounted to a mock fuselage of one-half of a 76 mm by 381 mm ellipsoid centered at mid-chord, which is used to simulate the flow near the vehicle body. The fuselage is mounted to a large 610 mm by 230 mm flat acrylic plate to mimic a symmetric boundary condition and thereby simulate the entire wing system. The water level in the tank is placed 13 mm above the acrylic plate in all tests.

The wing joints are actuated through counter-tensioned steel cables to an off-board bank of three EX-106+ Dynamixel motors, following trajectory commands over serial. These motors are mounted to a ATI Gamma 6-axis force transducer, which is towed down the tank. All data logging and serial communication is performed in LabVIEW, while MATLAB is used for the final data processing.

The force signatures of interest (namely unsteady lift, drag, added mass, and vortex lift) are relatively low frequency, on the order of the flapping frequency ($0.35$ Hz) or chordwise vortex shedding ($0.2U/c = 0.26$ Hz). Other known higher frequency force oscillations can be deemed unimportant; for example, the periodic breakup of the foil's thin drag wake at zero angle of attack. This frequency mismatch provides a theoretical basis for selecting the proper filtering parameters for rejecting sensor noise without losing data. The breakup frequency can be approximated from the wake width $w = cC_{D_\text{wing}}$ \cite{roshko1954drag} at zero angle of attack, as determined by the steady towed drag measurements in Fig. \ref{fig:StaticWing}. 
 \begin{equation}
 f_\text{breakup} \approxeq \frac{0.2U}{w} = 6\,\,\text{Hz} \label{eq:forcefilter}
 \end{equation}
 
We therefore low-pass filter the force data in MATLAB with a fifth order Butterworth filter with cutoff frequency equal to $f_\text{breakup}/2 = 3$ Hz (given the finite roll-off of the filter), while retaining slower frequencies.
 
Finally, the force felt by the transducer measures the sum of both the fluid force \emph{and} the force required to create the acceleration of the wing mass (inertia effects). We therefore run a rigid-body simulation in Featherstone's Spatial\_v2 MATLAB toolbox \cite{featherstone2008rigid} for the flooded body geometry in a vacuum with gravity and use the simulation result to correct the measured force.

\section{EXPERIMENT PARAMETERIZATION}
\subsection{Flapping Parameters}
As a standard within propeller theory and subsequently carried into 3D flapping research \cite{techet2008propulsive}, the location of 0.7 semispan from the flapping axis can be thought of as a representative slice for 2D analysis. We therefore base most of our parameterization off of this point, and subsequently analyze deviations due to 3D effects. 

As described in Sec. \ref{proto}, the semispan $s$ is three times the chordlength, and the spanwise location of the flapping joint $s_{2}$ is the same as the chord. The representative span $s_\text{rep}$ is therefore:
\begin{equation}
s_\text{rep} = s_{2}+0.7(s-s_2) = 2.4c \label{eq:repSpan}
\end{equation}
We assume that the flapping arclength $2h$ at the representative span is a good estimate of the wake width, and thereby nondimensionalize the flapping frequency with respect to the towing speed $U$ and flapping period $T$ as a Strouhal number:
\begin{equation}
\St = \frac{2h}{UT} \label{eq:Strouhal}
\end{equation}

In experimental flapping propulsion studies, high thrust efficiencies have been found at Strouhal frequencies $0.2\leq\St\leq0.4$ \cite{triantafyllou1991wake}. For this work, we will focus on the higher end of the flapping frequencies $\St=0.4$, as consistent with the results of \cite{izraelevitz2014adding,licht2010line} for better in-line motion performance. The $h/c$ ratio is set to $h/c=0.75$, constrained by the angular limits of the motor actuators and maintaining ease of comparison to prior 2D experiments \cite{izraelevitz2014adding}. This parameter combination reduces to a flapping frequency of $f = 0.35$ Hz.

The motions of the in-line and flapping joints are set as cosines, with $t=0$ as the beginning of the downstroke. The ratio of amplitudes is given by the stroke angle $\beta$;
\begin{equation}
A_1 = A_2/\tan(-\beta) \;\;\;\;\; A_2 = h/(s_\text{rep}-s_{2})\label{eq:A2}
\end{equation}
\begin{equation}
\theta_1(t) = A_1\cos(2\pi t/T) \;\;\;\;\; \theta_2(t) = A_2\cos(2\pi t/T) \label{eq:th12}
\end{equation}

Similar to the definition in \cite{tobalske1996flight}, $\beta<90^\circ$ is a bird-like trajectory with a forwards-traveling downstroke, and $\beta>90^\circ$ is a turtle-like trajectory with a backwards-traveling downstroke. The foil global velocity components $v_1$ and $v_2$ are defined in a frame coincident with the representative span but independent of foil rotation:
\begin{equation}
\Delta s_\text{rep}=s_\text{rep}-s_2 \label{eq:dels}
\end{equation}
\begin{equation}
v_1 = -\theta_1\dot{\theta}_2\Delta s_\text{rep}\sin(\theta_2)+\dot{\theta}_1[s_2+\Delta s_\text{rep} \cos(\theta_2)]+U\cos(\theta_1) \label{eq:vx}
\end{equation}
\begin{equation}
v_2 = \dot{\theta}_2\Delta s_\text{rep}-U\sin(\theta_1)\sin(\theta_2) \label{eq:vy}
\end{equation}

The pitching angle $\theta_3$ is defined by imposing a functional angle of attack of the foil directly, rather than the pitch angle, as shown to increase propulsive efficiency by \cite{hover2004effect}. We therefore compute the instantaneous angle of foil motion:
\begin{equation}
\theta_\text{motion}(t) = \arctan(v_2/v_1) \label{eq:AoF}
\end{equation}

Ignoring induced flow effects for the simplicity of the kinematic definition, we assume that the angle of flow is approximately the same as the angle of motion. We can thereby define an angle of attack at the rotation axis of 1/4 chord:
\begin{equation}
\alpha(t) = \theta_3(t) - \theta_\text{flow}(t) \approxeq \theta_3(t) - \theta_\text{motion}(t) \label{eq:AoA}
\end{equation}

Finally, as noted by \cite{theodorsen1935general}, the linearized angle of attack at the 3/4 chord point $\alpha_\text{eff}$ is the defining parameter, as opposed to the 1/4 chord point, to compensate for rotation-based lift:
\begin{equation}
\alpha_\text{eff}(t)= \alpha(t)+\frac{c}{2V(t)}\dot{\theta}_3(t)\label{eq:alpha_eff}
\end{equation}

Where $V(t) = \sqrt{v_1^2+v_2^2}$. We can now impose a functional form onto $\alpha_\text{eff}(t)$ in order to reasonably impose the instantaneous lift, and integrate to give a pitching trajectory $\theta_3(t)$:
\begin{equation}
\dot{\theta}_3(t) = \frac{2V(t)}{c}(\alpha_\text{eff}(t)+\theta_\text{flow}(t)-\theta_3(t)) \label{eq:difEQ}
\end{equation}

In summary, the joint motions $\theta_1(t)$, $\theta_2(t)$, and $\theta_3(t)$ are derived from four parameters ($\St=0.4$, $45^\circ\leq\beta\leq135^\circ$, $\Rey=30,000$, $h/c=0.75$) and one functional trajectory $\alpha_\text{eff}(t)$.

\subsection{Flow Angle Variation \label{sec:flowang}} 
\begin{figure}[t]
\begin{centering}
\includegraphics[scale=1.0]{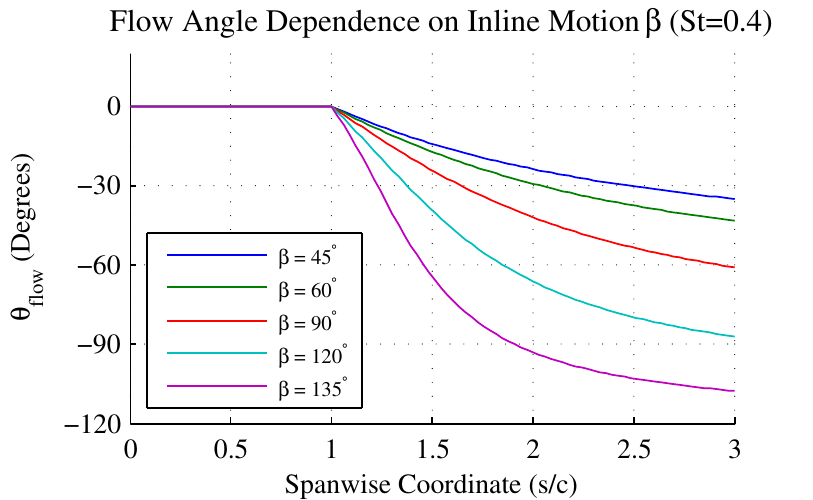}
\par\end{centering}
\caption{\emph{Flow Angle across Span} - Spanwise variation in the flow angle (middle of downstroke) is dependent on degree of in-line motion $\beta$ at fixed flapping frequency. \label{fig:FlowAngle}}
\end{figure}

The pitching angle $\theta_3$ is determined by the flow angle $\theta_\text{flow}$ at the representative span, but $\theta_\text{flow}$ can be a strong function of the spanwise coordinate. For example, Fig. \ref{fig:FlowAngle} illustrates the changing flow angle with respect to the degree of in-line motion $\beta$, at the same flapping frequency $\St=0.4$ in the middle of the downstroke. Note how the flow angle rapidly shifts to almost $\alpha_\text{max}=90^\circ$ over a short span for the turtle-like motion at $\beta=135^\circ$. However, the extent of variation is less rapid for symmetric flapping $\beta=90^\circ$, and can be even less rapid for smaller flapping frequencies.

Determining the pitch profile is therefore difficult, as the wing pitch must be kept close to $\theta_\text{flow}$ to avoid excessive local angle of attack. The spanwise pitching extent of the wing, in addition to the pitch angle itself, must be carefully chosen with respect to the wing trajectory. For example, flapping vehicles focusing on symmetric flapping will necessarily create a linear pitching profile over the majority of the wing, either actively \cite{send2012artificial} or passively \cite{kinkade2001ornithopter,lentink2010scalable,keennon2012development}. In the case of this experimental prototype, trajectories with strong in-line motion ($\beta \approxeq 135^\circ$) are the focus, requiring that the spanwise extent of pitching instead be small and the symmetric flapping capability is necessarily compromised. For bird-like trajectories ($\beta < 90^\circ$), the spanwise extent of pitching is less critical. The spanwise variation of the flow angle $\theta_\text{flow}$ is smaller, and only small wing-pitch is required to develop the necessary downstroke angle of attack.

\section{RESULTS}
\subsection{Static Wing Tests} 
\begin{figure}[t]
\begin{centering}
\includegraphics[scale=1.0]{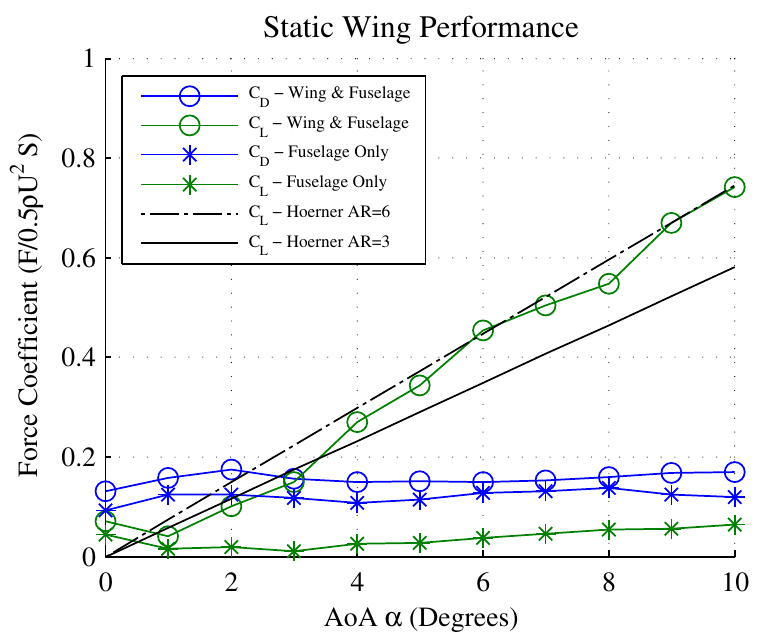}
\par\end{centering}
\caption{\emph{Static Wing Towing Tests} - Lift and drag for the stationary half-vehicle, with and without wing attached. Measurements averaged over three runs after steady state (3/4 tank length, or $Ut/c >= 6$), and compared to Hoerner's approximation \protect\cite{hoerner1985fluid} for finite aspect ratio airfoils. \label{fig:StaticWing}}
\end{figure}

The gliding performance of the half-vehicle model is illustrated in Fig. \ref{fig:StaticWing}. Lift and drag values have been normalized by the planform area $S=sc$ of the semi-span:
\begin{equation}
C_L = \frac{F_y}{0.5\rho SU^2 } \;\;\; C_D = \frac{-F_x}{0.5\rho SU^2} \;\;\; C_M = \frac{M_z}{0.5\rho ScU^2}\label{eq:CLCD}
\end{equation}

The lift data for the static wing agrees with Hoerner's results for finite aspect-ratio $\AR$ wings of rectangular planform \cite{hoerner1985fluid}:
\begin{equation}
\frac{dC_L}{d\alpha} = \Big(\frac{1}{1.8\pi} + \frac{1}{\pi \AR} + \frac{1}{2\pi(\AR)^2}\Big)^{-1} \label{eq:Hoerner}
\end{equation}

Induced flow effects, as expected, reduce the effective steady angle of attack - delaying stall and lowering the lift coefficient slope. In the case of this experiment, the semispan $\AR=3$, but an ideal symmetric boundary condition will double the effective $\AR$. The close match between the $\AR=6$ Hoerner approximation and the measured lift validates our implementation of the symmetric boundary in the experiment.

Note that the majority of the vehicle drag comes from the fuselage model, as expected from the unstreamlined supporting struts, large plate area, and wave drag from the free surface. However, the intent of the fuselage model is to create a realistic boundary condition for the flapping actuator, rather than mimic the force performance of the fuselage itself. The fuselage drag will therefore be subtracted from the subsequently presented actuator datasets in Sec. \ref{sec:sym} through Sec. \ref{sec:bird}, and optimizing the fuselage characteristics will be addressed in future work.

\subsection{Symmetric Flapping \label{sec:sym}} 
\begin{figure}[t]
\begin{centering}
\includegraphics[scale=1.0]{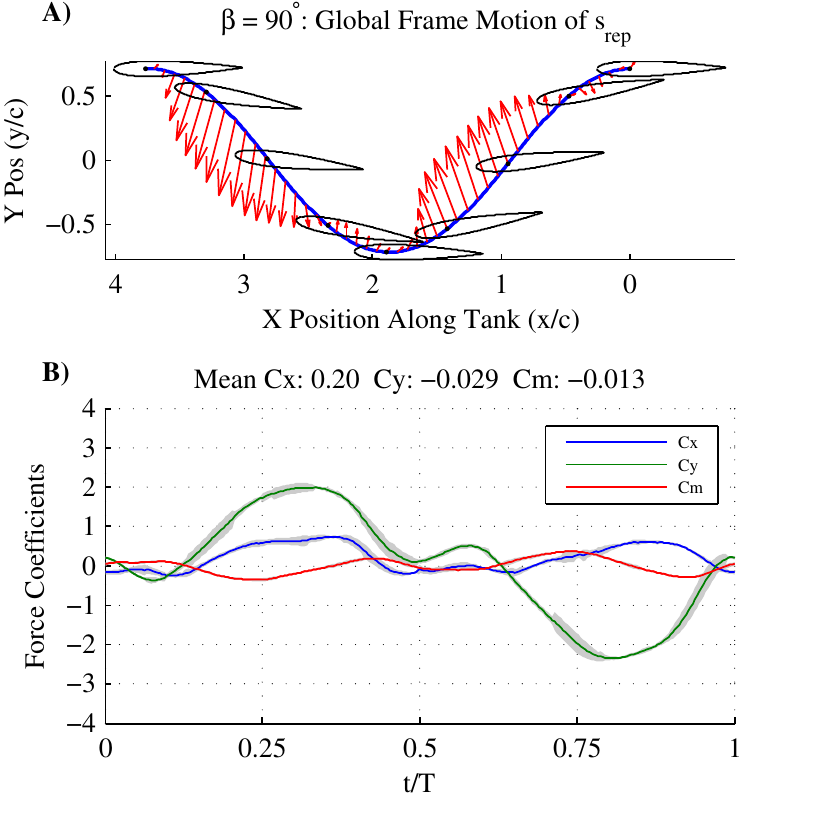}
\par\end{centering}
\caption{\emph{Symmetric Trajectory} - Symmetric trajectory with $\St=0.4$, $\beta=90^\circ$, $h/c=0.75$, $\alpha_\text{max}=45^\circ$. In upper plot (A), force coefficient vectors have been scaled by 1/4 with respect to the displayed chordlength and foils are plotted every $t/T=1/8$. Bottom plot (B) gives a time history of the force coefficients for the trajectory, illustrating the oscillating vertical force $C_y$ and two thrust $C_x$ peaks. Gray shading about force data gives three times the standard deviation over five runs, down-selected from eight total runs to remove spurious trials. \label{fig:SymStroke}}
\end{figure}

Symmetric trajectories, while not the focus of this research, provide a basis for comparison with traditional flapping foil experiments. Fig. \ref{fig:SymStroke} illustrates a symmetric trajectories ($\beta=90^\circ$) at the Strouhal frequency of $\St=0.4$ with a maximum angle of attack of the downstroke of $\alpha_\text{max}=45^\circ$. The angle of attack at $s_\text{rep}$ is imposed a functional form, in this case a sinusoid out of phase with the heaving motion. In only this symmetric case, given that $\dot{\theta}_3$ is small, we use the angle of attack $\alpha$ at 1/4 chord rather than 3/4 chord $\alpha_\text{eff}$ to define the motion to keep consistency with symmetric flapping foil studies \cite{hover2004effect}:
\begin{equation}
\alpha(t) = \alpha_\text{max}\sin(2\pi t/T) \label{eq:symmaoa}
\end{equation}

The pitch $\theta_3(t)$ is as derived in (\ref{eq:AoA}), and $\theta_1(t)$ and $\theta_2(t)$ as given by (\ref{eq:th12}). While $\alpha_\text{max}=45^\circ$ is higher than the steady stall angle of a NACA0013, the foil will not necessarily lose lift. In this trajectory, the foil quickly exceeds and leaves the stall angle at a rate faster than the stall dynamics \cite{dickinson1993unsteady}, similar to other flapping foil studies \cite{licht2010line,hover2004effect,techet2008propulsive}. 

For the purposes of distinction between the lift force, which is perpendicular to the time-varying flow angle $\theta_\text{flow}$, and the truly vertical force perpendicular to the freestream $U$, we now introduce new force coefficients aligned in a coordinate frame fixed to the vehicle fuselage:
\begin{equation}
C_x = \frac{F_x}{0.5\rho SU^2 } \;\;\;\; C_y = \frac{F_y}{0.5\rho SU^2 } \;\;\;\; C_m = \frac{M_z}{0.5\rho ScU^2} \label{eq:Cxym}
\end{equation}

The illustrated flapping profile generates a mean thrust of $C_x=0.20$ and negligible net vertical force $C_y=-0.029$ due to the trajectory symmetry. However, the instantaneous vertical force peaks to $\max|C_y(t)|\approxeq2$, dwarfing the thrust and would create a large heaving motion on the vehicle.

\subsection{Turtle-like Thrust Strokes \label{sec:turtle}}
\begin{figure}[t]
\begin{centering}
\includegraphics[scale=1.0]{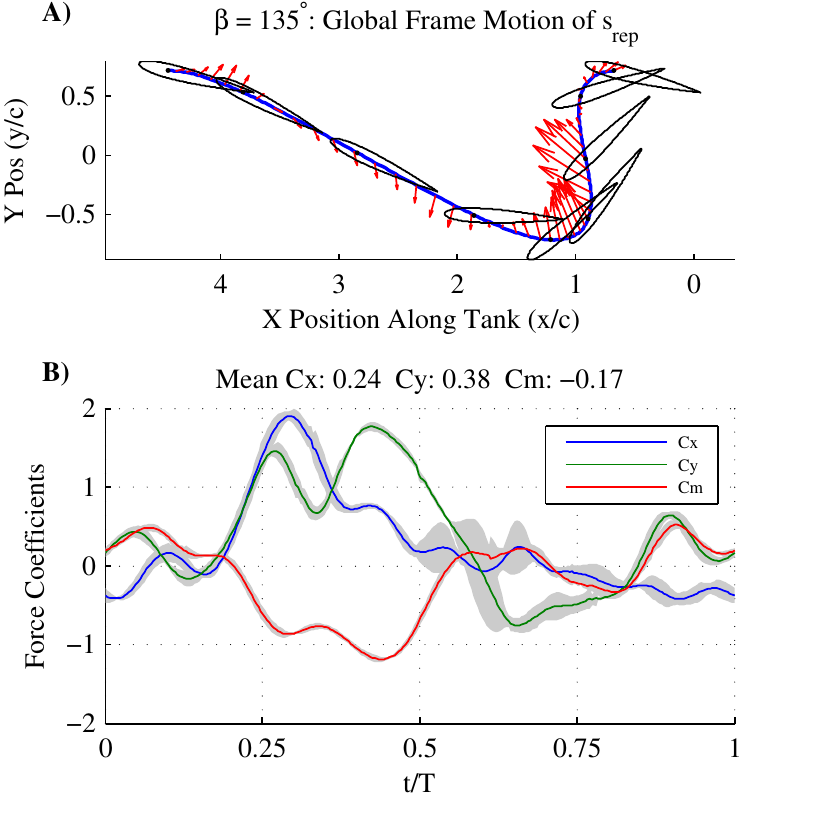}
\par\end{centering}
\caption{\emph{Turtle-like Trajectory} - Trajectory with $\St=0.4$, $\beta=135^\circ$, $h/c=0.75$, $\alpha_\text{max}=45^\circ$. Motion with respect to the global frame given in (A), and force history in (B). Similar to Fig. \protect\ref{fig:SymStroke}, gray region illustrates 3 standard deviation over 5 selected runs. \label{fig:TurtleStroke}}
\end{figure}

Using a backwards-traveling downstroke at stroke angle $\beta=135^\circ$ creates turtle-like flapping behavior, as indicated in Fig. \ref{fig:TurtleStroke}. The intent of this type of trajectory is to create thrust without oscillating vertical force, allowing the hypothetical vehicle to travel level underwater. The designed angle of attack profile is asymmetric, as the downstroke is intended to create most of the force, and no force during the upstroke.
\begin{equation}
\alpha_\text{eff}(t)= \Bigg \{ \begin{array}{cc}
\alpha_\text{max}(0.5-0.5\cos(4\pi ft)) & t \leqq T/2\\
0 & t>T/2
\end{array}\label{eq:downstrokeAoA}
\end{equation}

The flap illustrated in Fig. \ref{fig:TurtleStroke} includes the rotation correction, with $\alpha_\text{max}=45^\circ$, $\beta=135^\circ$, and $Str=0.4$. Note that the force on the foil is mostly centered during the downstroke, as intended, and creates a single thrust peak.

The rotation correction $\alpha_\text{eff}(t)$ vs $\alpha(t)$ works well at the beginning of the downstroke ($0<t/T<0.1$), mitigating most of the vertical force $C_y$. During the peak of the downstroke ($t/T=0.25$), the large $C_y$ is undesirable, but is unavoidable due to the limited lift to drag ratio of the foil. At the end of the downstroke ($t/T=0.4$), $C_y$ is large, but the foil is not yet rotating, indicating that this force is likely due instead to wake effects, such as a shedding LEV.

Optimized flapping trajectories, presented for the 2D case in \cite{izraelevitz2014adding}, are able to cancel this vertical force. We therefore expect, given a suitable optimization model, that intelligent modifications the trajectory would also improve the 3D case. As is, the current trajectory would force the vehicle in a combined vertical and horizontal motion, with little oscillation perpendicular to that path.

The net thrust force of $C_x=0.24$ is greater than the analogous symmetric case $C_x=0.20$ in Fig. \ref{fig:SymStroke}. The increased thrust is likely caused by a more constant angle of attack over the wingspan. The pitching profile at $\beta=135^\circ$ better matches the $\theta_\text{flow}(s)$ from Fig. \ref{fig:FlowAngle}, allowing the asymmetric $\beta=135^\circ$ trajectory to take advantage of more of the wingspan.

\subsection{Bird-like Lift Strokes \label{sec:bird}} 
\begin{figure}[tb]
\begin{centering}
\includegraphics[scale=1.0]{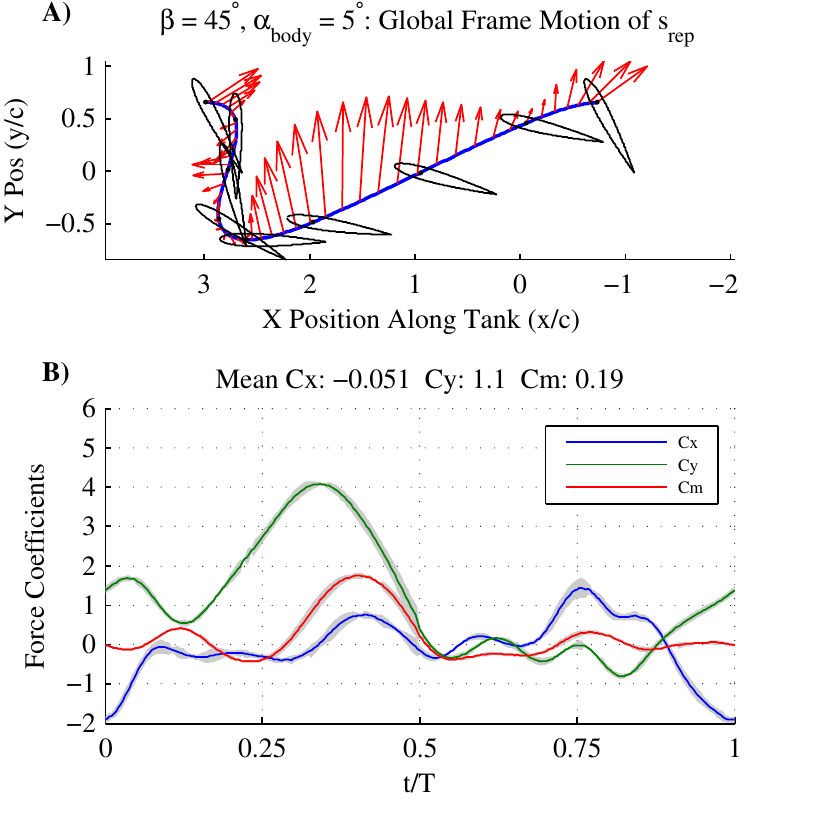}
\par\end{centering}
\caption{\emph{Bird-like Trajectory} - Trajectory with $\St=0.4$, $\beta=90^\circ$, $h/c=0.75$, $\alpha_\text{max}=45^\circ$. Motion with respect to the global frame given in (A), and force history in (B). Similar to Fig. \protect\ref{fig:SymStroke} and \protect\ref{fig:TurtleStroke}, the gray region illustrates 3 standard deviation over 5 selected runs.  \label{fig:BirdStroke}}
\end{figure}

Choosing a forwards-traveling downstroke $\beta=45^\circ$ demonstrates a bird-like flapping trajectory with boosted vertical force for aerial travel (Fig. \ref{fig:BirdStroke}). The flapping frequency remains $\St=0.4$, and choice of angle of attack profile is identical to (\ref{eq:difEQ}) to mitigate the rotational lift effects.

In this trajectory, we also set the fuselage angle of attack $\alpha_\text{body}=5^\circ$, to simulate the action of a hypothetical elevator, allowing the vehicle to take advantage of the entire wingspan.

The total average vertical force is $C_y=1.1$, over triple the gliding lift from Fig. \ref{fig:StaticWing}. This force comes at the cost of expended energy by the flapping actuator, so it does not increase the lift to drag ratio, but would decrease the stall speed of the aerial vehicle. While $C_y=1.1$ is not necessarily impressive in general for airfoil lift, it is large in this case given the uncambered foil and limited pitched area.

The symmetric flapping case (Fig \ref{fig:SymStroke}) makes a better comparison. If a hypothetical vehicle chose between the symmetric downstroke or bird-like downstroke for vertical force generation, the bird-like trajectory creates double the vertical force for the same $\alpha_\text{max}$ and flapping frequency. The symmetric case for $\alpha_\text{body}=5^\circ$ was not tested, so this is not a true comparison, but we expect the added steady lift from $\alpha_\text{body}$ to be small, given the static experiments (Fig \ref{fig:StaticWing}). The boosted vertical force is primarily due to the forwards travel of the downstroke, increasing the relative flow velocity.

The instantaneous force from the wing is not as smooth as the turtle-like trajectory (Fig. \ref{fig:TurtleStroke}). In this case, the rotational effects at the end of the downstroke are mitigated ($t/T=0.5$), and the upstroke thrust ($0.6<t/T<0.8$) is useful for aerial travel. However, the beginning of the downstroke ($t/T>0.8$ and $t/T<0.2$) has strong unintended drag, possibly due to the large angle of attack at the wing root. A slightly positive thrust force $C_x$, rather than the given nearzero thrust average, is required for aerial locomotion to overcome fuselage drag. Again, we expect further optimization of the pitching angle $\theta_3$ to mitigate the unwanted instantaneous forces.

\subsection{Hypothetical Vehicle Parameters}

This manuscript is a study in the actuation performance of a prototype wing design, to inform future vehicle parameter selection. For exact Reynolds $\Rey$ equivalence, the derived force coefficients are valid for water transport at $U=0.2$ m/s and air transport at $U=2.5$ m/s. For exact Strouhal $\St$ equivalence, the flapping frequency should also scale with the velocity: $f=0.35$ Hz in water and $f=4.4$ Hz in air. 

Given the force performance $C_x(\Rey,\St)$, $C_y(\Rey,\St)$, and $C_m(\Rey,\St)$ of each of the presented flapping trajectories at the given $\Rey$ and $\St$, we can now derive the sizing of a vehicle using this actuation method. For the horizontal locomotion, the steady state speed is given by the thrust force balance:
\begin{equation}
0.5\rho SU^2C_x = F_{x_{\text{wing}}} = F_{D_{\text{fuselage}}} = 0.5\rho AU^2C_D   \label{eq:Uwater}
\end{equation}

The turtle-like trajectory thrust coefficient of $C_x=0.24$ would therefore propel an underwater vehicle at speed $U=0.2$ m/s when $C_DA=C_xS=0.033$. Furthermore, we expect the experiment force coefficients are weakly Reynolds sensitive but strongly Strouhal sensitive, meaning that this derived $C_DA$ is also valid at close speeds ($3,000 < \Rey < 300,000$) as long as the flapping frequency $f$ also scales linearly with $U$ to maintain Strouhal $\St=0.4$.

Using the known drag coefficient of $C_{D_\text{wet}}=0.013$ on 5:1 streamlined axisymmetric fuselage for the experiment Reynolds number \cite{hoerner1965fluid}, the prototype wing could propel a large fuselage with wetted area of $A=2.6$ m\textsuperscript{2}. Clearly, the wing thrust is more than sufficient to propel a small 1 m long vehicle underwater at steady state and will be decreased on future wing designs. Similarly, vertical force $C_y=0.38$ from the turtle-like trajectory should also be reduced, but can be more easily offset by buoyancy or fixed lifting surfaces.

The vertical force for bird-like trajectories is larger, with average $C_y=1.1$. We can derive the vehicle flight mass by the vertical force balance:
\begin{equation}
0.5\rho SU^2C_y = mg   \label{eq:Uair}
\end{equation}

A vehicle moving at 2.5 m/s in air ($\rho=1.23$ kg/m\textsuperscript{3} \cite{hoerner1965fluid}) with this lift coefficient would have a weight budget of only $0.060$ kg. However, by again arguing a weak Reynolds sensitivity but strong Strouhal sensitivity, this lift coefficient also roughly applies at $U=7.5$ m/s if $f=13.2$ Hz, allowing a weight budget of $0.54$ kg. These flight weights and flapping frequencies are within reason for other flapping aerial vehicles \cite{grauer2009inertial,send2012artificial,wood2008first}. Additionally, the lift coefficient could likely be improved upon by subsequent iterations of the wing design. The 2D results predict upwards of $C_y=4$ for in-line motion trajectories, and the lower force coefficient for this wing is likely due to the flapping of only two-thirds of the planform.
\addtolength{\textheight}{-1.8cm}

\section{CONCLUSIONS}

In conclusion, the use of in-line motion for turtle-like thrust generation and bird-like lift generation has been fully realized in a 3D wing rotating from the wing root, extending the 2D pitching/heaving results of  \cite{izraelevitz2014adding}. The presented wing geometry has adequate force performance in both flapping regimes, and creates thrust and lifting trajectories that have stronger force than the analogous symmetric trajectory.

The current wing implementation uses unoptimized pitching trajectories $\theta_3(t)$, determined open-loop. We expect that the development of a model-based optimization loop would mitigate much of the instantaneous force unsteadiness. 

The spanwise variation in the flow angle $\theta_\text{flow}(s,t)$, as opposed to simply the flow angle at the representative span, is an important parameter that informs the wing geometry. A wing designed for symmetric flapping will have more distributed pitching than a wing designed for in-line motion.

For the given thrust and lift coefficients measured, we would predict actuation over-performance for an underwater vehicle, and adequate performance for aerial weight support. The experimental effectiveness of this wing definitively validates the use of in-line motion for boosting the force envelope of 3D flapping foils, with the specific application on the force envelope required for aerial/aquatic vehicles.

\section*{ACKNOWLEDGMENT}

We wish to thank M. Culpepper, R. Tedrake, D. Barrett, and the MIT Towing Tank laboratory for their thoughtful discussions on fluid mechanics, mechanical design, and experiment implementation. H. Guarino and K. Stephanopoulos aided the experiment setup, and A. Maertens, J. Dusek, L. Mendelson, and G. Bousquet contributed editing to the manuscript.

\bibliographystyle{IEEEtran_NoURL_YesBachelors}
\bibliography{root}

\end{document}